\documentclass[prl,twocolumn]{revtex4}
\usepackage{graphicx}
\renewcommand{\v}[1]{{\bf #1}}
\renewcommand{\b}[1]{\bar{#1}}

\def\eqa{\begin{eqnarray}}
\def\eea{\end{eqnarray}}
\newcommand{\eq}{\begin{equation}}
\newcommand{\ee}{\end{equation}}

\begin{document}

\title{Quantum Ginzburg-Landau theory of doped Mott insulators}
\author{Qiang-Hua Wang}
\affiliation{National Laboratory of Solid State Microstructures,
Institute for Solid State Physics, Nanjing University, Nanjing
210093, China}

%\date{\today}

\begin{abstract}
We improve a previous theory of doped Mott insulators with duality
between pairing and magnetism by a further duality transform. As
the result we obtained a quantum Ginzburg-Landau theory describing
the Cooper pair condensate and the dual of spin condensate. We
address the superconductivity by doping a Mott insulator, which we
call Mott superconductivity. Some fingerprints of such novelty in
cuprates are the scaling between neutron resonance energy and
superfluid density, and the induced {\it quantized} spin moment by
vortices or Zn impurity (together with circulating charge
supper-current to be checked by experiments).
\end{abstract}

 \maketitle

After more than a decade of intense experimental and theoretical
studies on high-$T_c$ superconductors, it is becoming clear that
the superconducting state turns out to be more ``normal'' than the
normal state in the sense that there are better-defined fermionic
quasi-particles that can be partly understood in terms of d-wave
Bardeen-Cooper-Schrieffer (BCS) theory subject to suitable
modifications. For example, the concept of nodal quasiparticles in
d-wave BCS theory seems to account for a handful of transport
properties.\cite{hightc} Recently it is even established by
scanning tunneling microscopy (STM) that the response of the
superconducting state to weak impurities is excellently described
by a usual d-wave Bogoliubov-de Genn (BdG)
theory,\cite{hoffman1karl} with quasi-particle scattering
interference (a signature of well-defined free quasi-particles).
The pairing gap function fitted according to scattering
interference idea\cite{qhwang} agrees unexpectedly with that from
angle-resolved photo-emission spectrum \cite{arpes} (which is a
probe in the momentum space in contrast to the real space STM).
The implications would be the superconducting state of high-$T_c$
superconductors is ``normal''. If this were the case, then what
makes high-$T_c$ superconducting state unique? The secrete is of
no secrete in the sense the experimental facts are already
existing. First, the effective charge carrier density is given by
doped holes. Second, the superfluid density scales with the doped
holes in the underdoped region instead of the electron density.
Third, there is well-defined inelastic neutron scattering
resonance around the magnetic vector $\v Q_0=(\pi,\pi)$ with
incommensurability that scales with the hole
density.\cite{resonance} The resonance energy scales with the
superfluid density. While the doping-dominated superfluid density
could be understood in one way or the other from re-normalized
Landau-Fermi liquid theory \cite{millis} or gauge
theory\cite{dunghai}, the existence of a neutron resonance is
unexpected, whose importance to seek a correct theory can never be
over-estimated, just as the isotope effect in usual
superconductors. The implication is twofold. The first is obvious:
There is dynamic spin anti-ferromagnetic correlations. The second
is more far-reaching. The resonance energy is related to the EM
photon gap in a Meissner state via Anderson-Higgs Mechanism. This
means that the condensate view spin excitations as if they were EM
gauge fields! This turns out to be just the crazy idea in a recent
theory of doped Mott insulators,\cite{crazy} which predicts a
topological mutual-duality between Cooper pair condensate and spin
magnetic condensate. The task of this Letter is to show that the
theory accounts for the above mentioned aspects of high-$T_c$
superconductivity. We identify the fingerprints of such
superconductors by a quantum Ginzburg-Landau theory of doped Mott
insulators, Eq.(\ref{newL}).

Previously \cite{crazy} we derived an exact ``all-boson''
representation of the electronic t-J model and identified the mean
field saddle point of the theory. The gauge fluctuations beyond
mean fields combined to slave particles were then integrated in
the infrared limit exactly, leaving us with an effective low
energy field theory of doped Mott insulators described by the
lagrangian $L=L_c+L_m+L_{CS}$ as follows.\cite{crazy}

First, $L_c$ describes the charge condensate (CC) of the theory,
\widetext \eqa & &L_c=i \b
\rho_c(\phi_c^*\frac{\partial_\tau}{i}\phi_c-2A_0)+iA_0
+i\delta\rho_c(\phi_c^*\frac{\partial_\tau}{i}\phi_c+2a^h_0-2A_0)
+i\v j_c\cdot(\phi_c^*\frac{\nabla}{i}\phi_c+2\v a^m-2\v
A)+\frac{u_c}{2}\delta\rho_c^2+\frac{1}{2K_c}\v
j_c^2,\label{lc}\eea where $\phi_c$ is the uni-modulus phase
factor of a spin-charge recombined Cooper pair, $\b\rho_c=(1-x)/2$
and $\v j_c$ are the Cooper pair density and spatial current,
respectively, $K_c$ is the effective zero-temperature superfluid
density, and finally $u_c$ is the effective local charge
repulsion. At low doping it can be shown that $K_c\propto x$. We
emphasize that $A_0$ in the first two terms of $L_c$ simply
reminds us that the effective charge density is $1-2\b \rho_c=x$,
but the Cooper pair density is $(1-x)/2$ (counting electrons) and
response to EM field as $2e$-carriers. This bears out the Mott
physics automatically. The effect of gauge fields $a^h_0$ and $\v
a^m$ will become clear below.

Second, $L_m$ describes the spin magnetic condensate (MC) of the
theory, \eqa & &L_m=i
\delta\rho_m(\phi_m^*\frac{\partial_\tau}{i}\phi_m-2a^m_0) +\v
j_m\cdot(\phi_m^*\frac{\nabla}{i}\phi_m-2\v
a^h)+\frac{u_m}{2}\delta\rho_m^2+\frac{1}{2K_m} \v
j_m^2,\label{lm}\eea
\endwidetext
where $\delta\rho_m$, $\phi_m$ and $\v j_m$ are the z-direction
magnetization $S^z$, the uni-modulus phase factor of $S^-$ and the
spin current. The background magnetic ordering wave vector $\v
Q=(\pi,\pi)(1-x)$ on square lattices as in cuprates. The effect of
gauge fields $a^m_0$ and $\v a^h$ will be clear below.

Third, $L_{CS}$ is given by \eqa L_{CS}&=&
\frac{i}{\pi}a^h_0(\nabla\times\v
a^h)_z+\frac{i}{\pi}a^m_0(\nabla\times\v a^m)_z.\label{lcs}\eea In
the above the subscript $z$ means the $z$-component of the object.
Let us observe that integration over $a^h_0$ and $a^m_0$ using the
total action $L$ enforces $\nabla\times \v
a^h=-2\pi\delta\rho_c\hat{z}$ and $\nabla\times \v
a^m=2\pi\delta\rho_m\hat{z}$. The physical meaning is that Cooper
pair condensate view localized $S^z$-moments as $2\pi$-flux
bundle, while the magnetic $S^-$ condensate view a localized
Cooper pair density fluctuation also as their $2\pi$-flux bundles.
(Note that locally $\delta\rho_m\leq 1/2$ and $\delta\rho_c\leq
1/2$ due to the microscopic no-double occupation constrain).
Similar ideas appear in the literature but suffer from U(1) gauge
symmetry breaking combined to slave particles.\cite{phasestring}

Except for the first two terms in $L_c$ (which gives the average
density of Cooper pairs and the charge carrier density) the two
condensates CC and MC are {\it mutually dual} to each other. This
mutual-duality is the crucial part of the theory that makes a
difference to the 1D counterpart in which spin-charge separation
is complete. In any case, the theory is millions of miles away
from a Landau-Fermi liquid.

At zero temperature, we can simplify the theory by applying the
concept of a self-duality transform as follows.\cite{dual} The
basic idea is, if a condensate, say MC is disordered, its vortices
condense instead. Let us call the new condensate as M$^\prime$C.
MC and M$^\prime$C are self-dual, and see the charges of each
other as flux bundles. Since $(\delta\rho_m,\v j_m)$ is a
conserved $3$-current due to the coupling to the phase mode
$\phi_m$, it can be represented by the field strength of a gauge
field, say, $b=(b_0,\v b)$ as \eqa
& &F_\mu=\epsilon_{\mu\nu\lambda}\partial_\nu b_\lambda,\\
& &\delta\rho_m=F_\tau/2\pi, \ \v j_m^{x,y}=\v F_{x,y}/2\pi. \eea
This gauge field is seen by M$^\prime$C via self-duality, and by
CC via mutual duality. These lines of reasoning suggests the
following rewriting of the total Lagrangian as \widetext \eqa
L=&&iA_0+i\b
\rho_c(\phi_c^*\frac{\partial_\tau}{i}\phi_c-2A_0)+i\delta\rho_c
(\phi_c^*\frac{\partial_\tau}{i}\phi_c-2A_0+2b_0)+i\v j_c \cdot
(\phi_c^*\frac{\nabla}{i}\phi_c-2\v A+2\v b)
+\frac{u_c}{2}\delta\rho_c^2+\frac{1}{2K_c}\v j_c^2\nonumber \\
&&+i\delta\rho_v(\phi_v^*\frac{\partial_\tau}{i} \phi_v-2b_0)+i\v
j_v\cdot (\phi_v^*\frac{\nabla}{i}\phi_v-2\v
b)+\frac{u_v}{2}\delta\rho_v^2+\frac{1}{2K_v}\v
j_v^2+\frac{u_m}{8\pi^2}F_\tau^2+\frac{1}{8\pi^2K_m}(F_x^2+F_y^2),\label{L}\eea\endwidetext
where $\phi_v$ is a uni-modulus phase factor describing the vortex
condensate M$^\prime$C and $j_v=(\delta\rho_v,\v j_v)$ is the
associated current, $u_v$ is the vortex fugacity and $K_v$ is the
vortex condensate phase stiffness. The Maxwell term for the
physical EM fields is implicit in the theory. A caution of gauge
symmetry related to the EM gauge fields $(A_0,\v A)$ is in order.
The Lagrangian density itself is not EM gauge invariant due to the
isolated term $iA_0$, but the total action $S=\int d^2xd\tau L$
is. The mutual-duality is now aptly described by the fact that
both CC and M$^\prime$C are coupled to the spin fluctuation gauge
field $b$. As a result the Chern-Simons term $L_{CS}$ can be
dropped from now on. $u_v$ and $K_v$ may be determined
microscopically (for example at the mean field level of the
bosonized t-J model\cite{crazy}), but their behavior follows from
simple arguments. Suppose we freeze the pairing channel, the
magnitude of $K_v$ is related to the gap for spin excitations.
This gap is zero in ordered MC at low dopings. At higher doping
levels, since the squeezed charges work as vortices to MC, they
eventually disorder the MC to its self-dual M$^\prime$C at a
critical doping level. At even higher doping, M$^\prime$C is
established with increasing phase stiffness. These lines of
reasoning suggests that the bare stiffness $K_v\sim K_c$. This is
more than reasonable in view of the fact that CC and M$^\prime$C
fight against a common animal, the spin excitation $b$-fields, and
can condense simultaneously to form a superconductor, for which we
propose a new terminology, a Mott superconductor, to discriminate
from the usual BCS superconductor. In any case, we just have to
tune a finite number of parameters for a specific material, but
the basic physics is captured by the theoretical framework.

The final form of the theory is obtained by integrating out the
auxiliary currents $(\delta\rho_c,\v j_c)$ and $(\delta\rho_v,\v
j_v)$ as \widetext \eqa L=&&iA_0+i\b
\rho_c(\phi_c^*\frac{\partial_\tau}{i}\phi_c-2A_0)+
\frac{1}{2u_c}(\phi_c^*\frac{\partial_\tau}{i}\phi_c-2A_0+2b_0)^2
+\frac{K_c}{2}(\phi_c^*\frac{\nabla}{i}\phi_c-2\v A+2\v b)^2\nonumber \\
&&+\frac{1}{2u_v}(\phi_v^*\frac{\partial_\tau}{i}
\phi_v-2b_0)^2+\frac{K_v}{2}(\phi_v^*\frac{\nabla}{i}\phi_v-2\v
b)^2+\frac{u_m}{8\pi^2}F_\tau^2+\frac{1}{8\pi^2K_m}(F_x^2+F_y^2),\label{newL}\eea\endwidetext
which is a quantum Ginzburg-Landau theory for doped Mott
insulators describing the Cooper pair condensate and the dual of
spin condensate. In applying to finite temperatures it is no
longer valid to do self-duality transform. This is because entropy
effect at finite temperatures requires a concise meaning of
order/disorder. In a sense temperature breaks the self-duality
symmetry. But there is no problem in applying the theory to finite
temperatures by keeping the above form.

A few lines of discussion on mechanism of Mott superconductor is
in order. In the literature it is hotly debated whether spin
fluctuations are the gluing force of pairing. The answer from the
present theory is NO. They do exist but act to compete with
superconductivity in a topological fashion. We would claim that
both pairing and magnetism are intrinsic properties of doped Mott
insulators.

The theory in the above new format place a good framework to
discuss the Mott superconductor. Some unique features of
Mott-superconductivity are immediately clear from this theory as
we now discuss.

1) The CC+M$^\prime$C state is a simultaneous Meissner state to
$A$-photons and $b$-photons. At zero energy they can only appear
in vortices of the condensate. As such, the gauge flux $\int d^2x
\nabla\times 2\v b=2\pi n$ is quantized, just as the flux
quantization in usual superconductors. The most-likely case is
each vortex carries $n=1$ quantum of flux, and according to
$\delta\rho_m=\nabla\times\v b/2\pi$ this is a quantized
spin-$1/2$ moment. This tendency was also observed in a bosonic
resonant valence bond theory based on the phase string effect on
bipartite lattices.\cite{phasestring} The length scale of the
distribution of this moment is determined by the ``London
penetration depth'' \eq
\lambda_b=\sqrt{\frac{u_m}{4\pi^2(K_c+K_v)}},\ee which has the
same qualitative behavior as that of the EM London penetration
depth $\lambda_L\sim 1/\sqrt{K_c}$, assuming $K_c\sim K_v$. Since
the dual of M$^\prime$C is MC, the vortex favors MC. In fact in
the vortex core, the gap for dynamic $b$-fluctuations is lowered.
Therefore, in addition to the quantized spin moment, we also
expect a cloud of softened spin excitations on an incommensurate
anti-ferromagnetic background (according to the starting point of
the theory). {\em The quantized spin moment is a benchmark to
compare with other theories \cite{so5} that also predict induced
magnetism}. This should also be a fingerprint of Mott
superconductivity.

In the absence of applied magnetic field $\nabla\times \v A=0$,
the vortex can be induced, for example, around Zn impurities in
cuprates. This is because the Zn impurity works as an empty site
to the system where unpaired spins have lower energy than
elsewhere. (A Ni impurity does not upset spin pairing and is
therefore a much weaker impurity than Zn.) The fact that Zn
impurity captures spin moment has been observed by nuclear
magnetic resonance (NMR) measurements.\cite{zinc} The next
challenge to both our theory and experimentalists is to prove that
there is also a charge super-current around Zn impurities, since
the gauge flux of $\v b$ is also seen by CC. In some sense this is
indirectly proved by the strong suppression of Zn impurity to
superconductivity, in a Swiss-cheese fashion\cite{zinc} as if each
Zn kills a super-conducting region of the order of $\pi\xi^2$, the
area of a vortex core.

On the other hand, under an applied magnetic field, two
possibilities follows. If the vortex core energy in M$^\prime$C is
lower than that in CC, then a quantized spin moment is induced in
the CC-vortex to counteract the applied flux. This lowers the
kinetic energy cost in CC, making the CC-vortex non-topological
(with zero winding number), with the expense of causing a vortex
in M$^\prime$C. This is a way of making cheap vortices as realized
in another context \cite{palee} (but without the topological
structure in M$^\prime$C). Moreover, the softened spin
fluctuations will be reflected in inelastic neutron scattering,
with the scattering strength scaling with the number of vortices.
This is exactly what was observed in cuprates. \cite{lake} Another
possibility is that the M$^\prime$C-vortex is more expensive than
CC-vortex. In this case the former is no longer inducible by the
latter, and one has the normal CC-vortex. It is very likely that
the first possibility is always the case in cuprates. The argument
is as follows. If the second possibility were true, there would be
a zero-energy peak in the single-particle density of states in the
vortex core of a BCS d-wave superconductor, which is actually
never observed in cuprates. \cite{corestm} This absence is however
compatible with the capturing of spin moment around the vortex,
making it non-topological. The existence of local spin moment and
the removal of super current (caused by phase winding) are
detrimental to the zero energy single particle state.

The underlying softened incommensurate anti-ferromagnetic modes
may be related to the checker board pattern observed in the local
density of states around vortices in cuprates.\cite{hoffman2} The
period of $4a$ in the modulation may be a result of
commensurability pinned by the underlying lattice.

2) The dynamic spin fluctuations are gapped. For the transverse
$b$-photon, the energy gap is given by $w_b\sim
\sqrt{K_m(K_v+K_c)}\sim 1/\lambda_b\sim 1/\lambda_L$ (assuming
$K_v\sim K_c$), tracing the superfluid density $1/\lambda_L^2$. In
view of our starting spin magnetic ordering vector $\v Q=\v
Q_0(1-x)$ for cuprates this is just the incommensurate inelastic
neutron resonance observed in cuprates.\cite{resonance} When the
$b$-photon wave-length decreases so that the spin excitations are
closer to the vector $\v Q_0$, the gap increases. These results
explain the existence of a $\v Q_0$ resonance at higher
energy.\cite{resonance} The academic importance of the fact that
the neutron resonance energy traces the superfluid density can
never be over-estimated: The superfluid density is the quantity
describing the response of the superconductor to physical EM
fields that is not of spin-origin in a usual BCS superconductor.
The relation between neutron resonance and superfluid density
therefore implies a breathtaking effect: the superfluid response
to spin excitations as if they were EM photons! This is just the
consequence of the mutual-duality in the theory. In this sense,
neutron resonance is just a fingerprint of a Mott superconductor,
in the same way that the isotope effect is the fingerprint of
phonon-driven weak-coupling BCS superconductors.

In the literature, there is a debate as whether the resonance is
some collective mode that drives superconducting transition. The
theory shows that it is merely a consequences rather than the
cause.

3) By inspection of $L$ in Eq.(\ref{L}) we find that assuming
absence of long-range forces between CC-charges and between
M$^\prime$C charges, the massless phase modes in the CC and
M$^\prime$C, combining with the gauge field $b$, couples to the CC
and M$^\prime$C $3$-currents. This is where one can do
re-fermionization by expressing these currents in terms of Dirac
fermions. The theory for such fermions has to assume the form of a
QED$_3$ theory by Lorenz invariance and gauge invariance. This is
indeed found as a property of nodal quasi-particles in d-wave
superconductors coupled to fluctuating vortices.\cite{qed3} Our
theory gives physical meaning to the $b$-gauge field. While our
theory starts off Mott insulator limit on microscopic grounds, the
QED$_3$ starts off a phenomenological description of
quasi-particles interacting with vortices in d-wave
superconductor. At zero doping our theory describes quantum
antiferromagnetism while QED$_3$ arrives at the same place via
chiral symmetry breaking of the field theory combined with
hand-made parameters to cope with the Mott insulator limit. Under
these contexts they address the same physics from different paths
that all lead to a place called Rome. The existence of low lying
fermion excitations is important to understand the well-behaved
quasi-particle scattering interference from weak
impurities,\cite{hoffman1karl} in that in the absence of vortices
in both CC and M$^\prime$C, the effective theory is very much the
same as that of a usual BCS superconductor as viewed in the CC
sector alone. The implication is that local density of states
(LDOS) modulation around weak impurities observed by scanning
tunnelling microscopy\cite{hoffman1karl} are the effect of
scattering interference of the low-lying quasi-particle states as
if in usual d-wave BCS theory, but the LDOS modulation found
around vortices \cite{hoffman2} finds another likely mechanism in
our theory, i.e., via capturing of quantized spin moments and
softening of incommensurate anti-ferromagnetic fluctuations.

\acknowledgements{I thank Z. Tesanovic for communications. This
work is supported by NSFC 10204011 and 10021001.}

\end{document}